\documentclass[journal]{IEEEtran}
\usepackage{graphicx}
\usepackage{amstext}
\usepackage{algorithm}
\usepackage{algorithmic}
\usepackage{mathrsfs}
\usepackage{amssymb}
\usepackage{amsmath}
\usepackage{bm}
\allowdisplaybreaks[4]
\usepackage{epstopdf}
\usepackage{multicol}
\usepackage{stfloats}
\usepackage{url}
\usepackage{color}
\usepackage{enumerate}
\usepackage{breqn}
\usepackage{bbm}
\usepackage{cite}
\usepackage[font=small]{caption}
\usepackage{subcaption}
\usepackage{enumitem}
\usepackage{array}
\usepackage{comment}

\usepackage{pifont}
\newcommand{\xmark}{\ding{55}}

\ifCLASSINFOpdf
\else
\fi
\begin{document}

\title{Secure Targeted Message Dissemination in IoT Using Blockchain Enabled Edge Computing}

\author{Muhammad Baqer Mollah,~\IEEEmembership{Student Member,~IEEE,}
        Md Abul Kalam Azad,~\IEEEmembership{Senior Member,~IEEE,}
        \\Yinghui Zhang,~\IEEEmembership{Member,~IEEE}%

\thanks{Muhammad Baqer Mollah and Md Abul Kalam Azad are with the Department of Computer Science and Engineering, Jahangirnagar University, Dhaka 1342, Bangladesh. (Emails: m.m.baqer@ieee.org; makazad@juniv.edu).}
\thanks{Yinghui Zhang is with the School of Cyberspace Security, Xi\textquotesingle an University of Posts and Telecommunications, Xi\textquotesingle an 710121, China. (Email: yhzhaang@163.com).}
\thanks{(Corresponding author: Yinghui Zhang and Md Abul Kalam Azad)}
\thanks{This research is partly supported by the National Natural Science Foundation of China under the grant no. 62072369 and the Youth Innovation Team of Shaanxi Universities under the grant no. 23JP160.}
\thanks{This article has been accepted for publication in IEEE Transactions on Consumer Electronics. This is the author's version which has not been fully edited and content may change prior to final publication.}
\thanks{DOI: https://doi.org/10.1109/TCE.2024.3436825}
\thanks{\copyright 2024 IEEE. Personal use is permitted, but republication/redistribution requires IEEE permission. See https://www.ieee.org/publications/rights/index.html for more information.}%
}


\markboth{IEEE Transactions on Consumer Electronics}%
{Shell \MakeLowercase{\textit{et al.}}: Bare Demo of IEEEtran.cls for IEEE Journals}

\maketitle

\begin{abstract}
    Smart devices are considered as an integral part of Internet of Things (IoT), have an aim to make a dynamic network to exchange information, collect data, analysis, and make optimal decisions in an autonomous way to achieve more efficient, automatic, and economical services. Message dissemination among these smart devices allows adding new features, sending updated instructions, alerts or safety messages, informing the pricing information or billing amount, incentives, and installing security patches. On one hand, such message disseminations are directly beneficial to the all parties involved in the IoT system. On the other hand, due to remote procedure, smart devices, vendors, and other involved authorities might have to meet a number of security, privacy, and performance related concerns while disseminating messages among targeted devices. To this end, in this paper, we design STarEdgeChain, a security and privacy aware targeted message dissemination in IoT to show how blockchain along with advanced cryptographic techniques are devoted to address such concerns. In fact, the STarEdgeChain employs a permissioned blockchain assisted edge computing in order to expedite a single signcrypted message dissemination among targeted groups of devices, at the same time avoiding the dependency of utilizing multiple unicasting approaches. Finally, we develop a software prototype of STarEdgeChain and show it's practicability for smart devices.
\end{abstract}


\IEEEpeerreviewmaketitle

\section{Introduction}
    \IEEEPARstart{T}{he} Internet of Things (IoT) \cite{vaezi2022cellular, 9509294, mollah2017secure}, has been rapidly growing more than ever before. From consumer electronic devices to industry, the IoT is making it possible to connect a variety of things and people to the Internet, and each other to bring enhancements to our quality of lives, surroundings, and system performances. Such things, also called smart devices (SDs) connect to the Internet and each other to exchange information, collect data, analyze, and make better decisions in an autonomous way without human operation. These abilities make SDs very popular in various application areas in the home, smart power grid, building, city, industry, healthcare, and many more. In specific, personal and home devices connected with IoT \cite{10384441, kim2020research} have been receiving much attention in the consumer electronics landscape since it is expected that the IoT is one of the key enablers to bring simplicity and ubiquitous connectivity to the consumers. 
    
    The increasing number of SDs is potentially demanding to add new features as well as receive latest instructions, safety messages, alerts, price information, and security patches from the service providers or respective authorities. Once the SDs are deployed, sending such messages is becoming an important practice. Moreover, the timely and targeted dissemination of such messages is also crucial. For instance, once any new bug or zero-day security vulnerability is revealed, the SDs become open to the potential attackers which may lead to attacks on other parts of the IoT system.
    
    Unfortunately, along with the usage of remote dissemination of messages to SDs necessarily has many benefits, it may raise a number of critical security and privacy related concerns against all the involved parties within the process, at the same time, efficient and timely distribution among the SDs. First concern is about privacy of SDs \cite{yang2022privacy, daidone2021blockchain, tan2021rsu, kim2020toward}. Intercepting the sensitive information of SDs related to identity and any other associated details may help malicious adversaries to identify the SDs as well as possible to launch attacks to the SDs by exploiting the vulnerabilities once published. Second concern is the confidentiality break of messages \cite{9102441, bauwens2020over, prathiba2021sdn}, where an adversary is able to eavesdrop the communication channel or attacks the repository to obtain the message, and afterwards does analysis for further launching attacks. Thirdly, hijacking to initiate replay an old but legitimate message and replacing a malicious code, instructions, or alerts with the original message on a stored server, cache repository or while transferring can be considered as integrity attacks. Finally, remote dissemination of messages could also suffer from efficient distribution and denial of service (DoS) \cite{zou2021distributed, gupta2021blockchain} or single point of failure (get compromised) \cite{ayaz2020proof, arul2021multi} type availability issues, where typical client-server based approaches become bottlenecks to distribute messages among a large number of SDs concurrently.
    
    Although cache-enabled repository \cite{ambrosin2014updaticator} can be used for ensuring the efficient targeted distribution, unfortunately, still the repositories may be targeted for availability and privacy types of attacks as well as have significant scalability barrier. In particular, in terms of privacy concerns, the cached distributed servers have knowledge which SD is asking for what, the adversary may monitor to obtain this. Meanwhile, edge computing is introduced where the cloud computing’s storage, network, and computational capabilities are extended to the network edge. In particular, in the context of edge computing based IoT \cite{datta2024blockchain, hong2024secure, li2023integrated, li2024blockchain, wang2024blockchain, wang2023blockchain}, edge computing nodes are placed closer to the end devices so that local data processing, storage, and cache repository as well as low latency like services can be offered. However, when edge computing node will consider as storage repository, it will raise a number concerns on edge security and privacy, such as how to secure the nodes, how to make the nodes trustworthy among the smart devices, how to maintain security for scalable nodes, and how to build storage repository among heterogeneous nodes. In this context, blockchain maintained by the edge nodes can be a suitable candidate for IoT-like scenarios. Conceptually, blockchain is referred to as a variety of distributed ledger technology, initially introduced in digital currency \cite{nakamoto2019bitcoin}, has some excellent features such as decentralization, security, time-stamped, temper-resistance, and trust.

    \textit{Motivation and Contribution:} Motivated by the aforesaid concerns, in this paper, we develop STarEdgeChain, a secure targeted message dissemination solution for IoT, which might have push or pull features using blockchain enabled edge computing. To the best of our knowledge, the developed architecture is the first work integrating targeted message dissemination with permissioned blockchain and edge computing. To sum up, the following can be a list of the key contributions of this research.
    
    \begin{itemize}
    \item We define an adversary model for the edge computing assisted message dissemination solution, where edge computing nodes are introduced as storage repositories to disseminate messages among the targeted smart devices. In this adversary model, the edge nodes are considered as semi-trusted (honest, but curious) nodes, and they are located in different places as distributed matter.
    
    \item Based on the adversary model, we introduce STarEdgeChain, which aims to provide privacy protection, authentication, confidentiality, integrity, access control, freshness, availability, and traceability.
    
    \item We employ a cipher-policy attribute based signcryption scheme to achieve anonymous authentication, confidentiality as well as fine-grained access control. Integrating with a trusted authority, tracking real identities in case of any malicious activities from insider entities can be possible.
    
    \item We develop a permissioned blockchain maintained by edge nodes and vendors/authorities into the STarEdgeChain to record the messages in immutable and verifiable blockchain to ensure strong message integrity, freshness, and availability.

    \item We present security and privacy analysis of developed STarEdgeChain. We also implement STarEdgeChain as a prototype to analyze the performance and practicability in terms of computational costs.
    
    \end{itemize}
    
    \textit{Paper Organization:} The remainder of this paper can be summarized as follows. Section II describes the problem statement including system settings, adversary model, and security requirements. The following section, the preliminary background of our construction is considered. Next, section IV presents the detailed construction of the proposed STarEdgeChain. After that, section V summarizes the security and performance analysis. Section VI discusses the related works. Finally, section VII concludes the presented work. The list of notations and their definitions used in this paper is described in Table \ref{tab: notations}.

\begin{table}[ht]
    \centering
    \caption{Summary of major notations and definitions.}
    \label{tab: notations}
    \begin{tabular}{m{1.6cm}|m{4.5cm}}
    
    \hline \hline

    \textbf{Notation}   &   \textbf{Definition} \\
    \hline

    $\lambda$    &    Security parameter \\
    \hline

    $p$    &    Prime number \\
    \hline

    $\mathbb{G}_1, \mathbb{G}_2, \mathbb{G}_3$    &    Three cyclic groups \\
    \hline

    $e$    &    Bilinear map \\
    \hline

    $g_1, g_2$    &    Generator of $\mathbb{G}_1$, generator of $\mathbb{G}_2$ \\
    \hline

    $\mathbb{Z}_p$    &    Integers modulo $p$ \\
    \hline

    $pk$    &    Public key \\
    \hline

    $sk$    &    Asymmetric secret key \\
    \hline

    $mk$    &    Master key \\
    \hline

    $key_{sym}$    &    Symmetric secret key \\
    \hline

    $key_{sign}$    &    Signing key \\
    \hline

    $key_{ver}$    &    Verification key \\
    \hline

    $msg$    &    Plaintext message \\
    \hline

    $\mathcal{S}$    &    The set of attributes \\
    \hline

    $\mathcal{T}$    &    Access tree structure \\
    \hline
    
    $\mathcal{A}$    &    An adversary \\
    \hline

    $\mathcal{ST}$    &   Signed Ciphertext \\
    \hline

    $\mathcal{CT}_{msg}$    &   Ciphertext of message \\
    \hline

    $\{0, 1\}^*$    &   Set of binary strings \\
    \hline

    $\mathcal{Y}$    &    Set of leaf nodes in access tree \\
    \hline

    $\mathcal{H}_1, \mathcal{H}_2$    &    Hash functions \\
    \hline

    $\pi$    &    Signature \\
    \hline

    $\Vert$    &    Concatenation operator \\
    \hline

    $\oplus$    &    XOR operation \\
    \hline

    \end{tabular}
\end{table}

\section{System Description and Design Goals}
    In this section, we first formalize the considered IoT system model, and then, outline the adversary model along with security and privacy properties of our proposed architecture.

\subsection{System Settings and Model}
\begin{figure} [!t]
	\includegraphics[width=\linewidth]{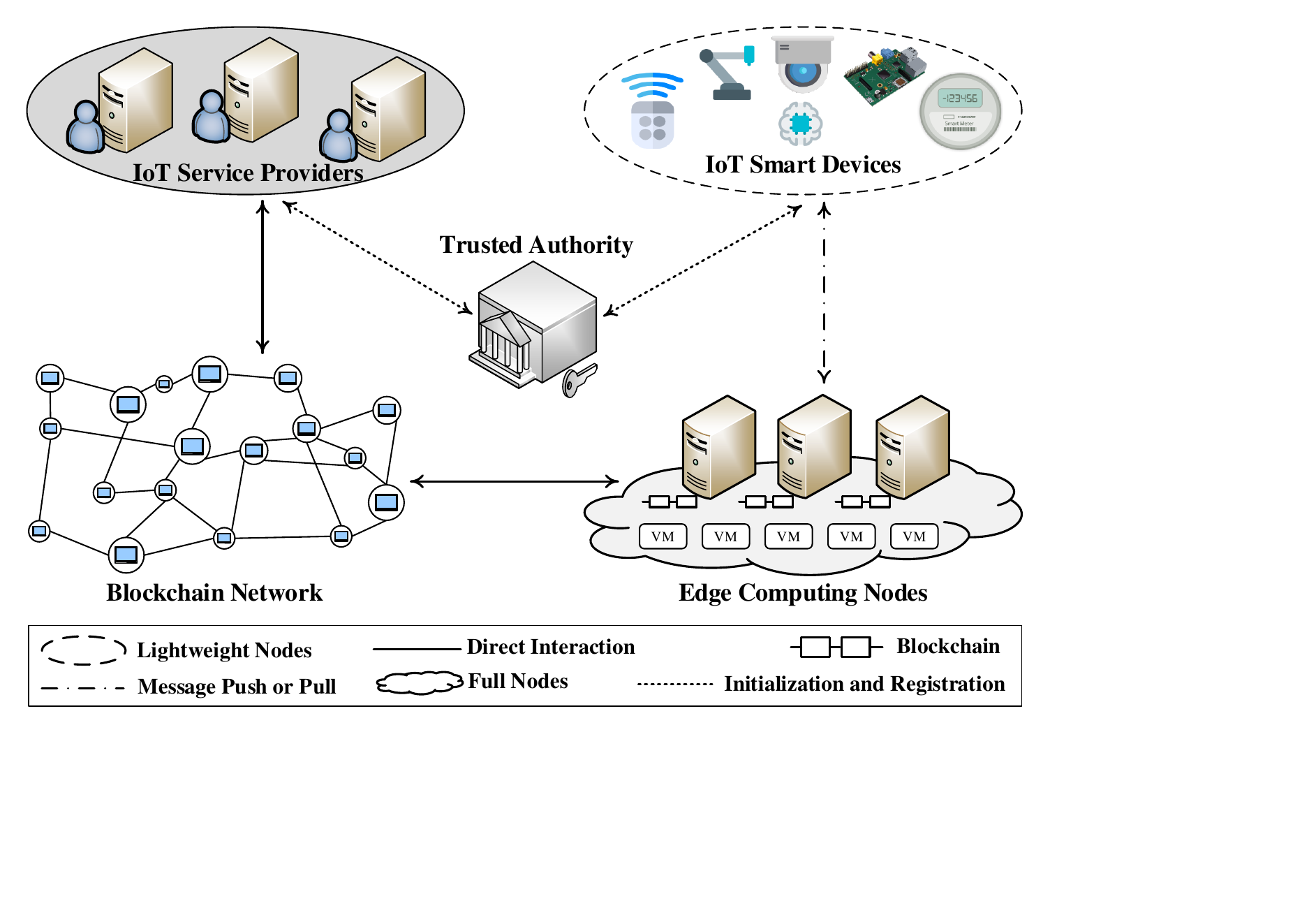}
	\caption{The considered IoT system model.}
    \label{fig: system}
\end{figure}

    The participants involved in STarEdgeChain are given below as shown in Fig. \ref{fig: system}.
    
    \begin{itemize}
    \item \textbf{Smart Devices (SD):} We consider generic SDs which contain operating systems and other application software to support IoT systems. The IoT system may be autonomous or semi-autonomous, but we assume that the SDs are remotely placed and maintained. However, due to limited computing and storage capabilities, the SDs can’t validate and store the entire blockchain, rather SDs are able to run cryptographic algorithms, and depended on receiving messages, such as new features, scheduled maintenance times, alerts, instructions, safety messages, advertisements, price information, security updates, etc. from the blockchain network.
    
    \item \textbf{Edge Computing Devices (ED):} The access points, gateways, aggregators, local clouds or fog nodes are typically referred as EDs that are deployed in between the cloud and SDs, i.e., close to the SDs by trusted agencies such as IoT authority, vendors or third party service providers. The EDs are assumed as semi-trusted nodes, i.e., as honest, but curious models. Moreover, these nodes can be considered as storage repositories or cache repositories, which are located in different places in a distributed way. We also assume that besides usual responsibilities, the EDs have capabilities to maintain the blockchain like storing the full copy of the blockchain.
    
    \item \textbf{Service Providers (SP):} The service providers may be owned by SDs vendors, IoT authority, control center or any other third party application software service providers. The responsibilities of SPs are to publish the message which they would like to disseminate among targeted SDs, send the message to the blockchain network for validation, and validate new blocks. 
    
    \item \textbf{Trusted Authority (TA):} TA is a centralized center which is mainly responsible for initializing the entities of the network as well as registration of all other entities. During registration, it generates cryptographic keys and delivers them securely to the entities. It may remain offline after initialization and key generation.
    \end{itemize}

\subsection{Adversary Model and Security Requirements}
    The communication among the SPs, EDs, and SDs is over a public channel. Hence, we consider the Delev-Yao threat model \cite{dolev1983security} in our work, where adversaries have full control over the communications channel, and are able to eavesdrop the messages and also, perform different attacks such as inject new illegitimate messages, replay and change messages, or spoof other identities. Moreover, we also assume that the TA is a  well-protected unit and trustworthy.
    
    However, to prevent the aforesaid malicious actions, our proposed STarEdgeChain will be considered to fulfill the following security and privacy properties:
    
    \begin{itemize}
    \item \textbf{Privacy Protection:} assures the anonymity of all involved parties and message requests so that any attacker cannot determine the real identity of any SD, and message information by doing analysis of messages, messages request, and blocks. Also, an adversary can’t find out which SD is asking for what messages.
    
    \item \textbf{Authentication:} ensures security from any illegitimate entities to access messages, and the SDs \& Blockchain validators can verify the authenticity of the messages, i.e., really came from the valid source or validated by trusted parties (for SDs).
    
    \item \textbf{Confidentiality:} refers to whether only the targeted SDs and validators that the vendors target can reveal it.

    \item \textbf{Integrity:} prevents adversaries from modifying or injecting the messages with malicious ones, and makes sure to verify the integrity of messages by all devices.
    
    \item \textbf{Fine-grain Access Control:} grants messages access on a fine-grained basis, i.e., revising the access attributes, setting the corresponding policies by any authority or vendor for their SD.
    
    \item \textbf{Freshness:} confirms that the receivers can verify that the messages are recent, and the adversary didn’t replay any old messages or make force to receive old messages.
    
    \item \textbf{Availability:} ensures high availability and resilience against DoS/DDoS/SPoF attacks of message dissemination compared to presently available client-server based architectures.

    \item \textbf{Traceability:} enables to track the identities of the involved entities so that it can be identified conveniently for any kind of dispute or misbehaving entities.
    \end{itemize}
    
\section{Preliminaries and Definitions}
    In this section, we introduce the preliminaries of our proposed STarEdgeChain. The preliminaries include two aspects. We first describe the definitions of bilinear pairing, decisional bilinear Diffie-Hellman assumption, and Cipher Policy-Attribute-Based SignCryption. Then, we present permissioned blockchain concept and consensus mechanism.

\subsection{Bilinear Pairings}
    \textbf{Definition 1 (Bilinear Map).} Let $\mathbb{G}_1$, $\mathbb{G}_2$, and $\mathbb{G}_3$ be three cyclic groups of the prime order $p$. Let $g_1$ and $g_2$ be the generators of $\mathbb{G}_1$ and $\mathbb{G}_2$, respectively, and $e$ denote a bilinear map, $e : \mathbb{G}_1 \times \mathbb{G}_2 \rightarrow \mathbb{G}_3$. Our proposed architecture makes utilize of the bilinear map $e$ which has the following three properties.
    
    \begin{itemize}
        \item \textit{Bilinear:} For all $g_1 \in \mathbb{G}_1$, $g_2 \in \mathbb{G}_2$ and $a, b \in \mathbb{Z}_p$, we have $e(g_1^a, g_2^b) = e(g_1, g_2)^{ab}$.
        \item \textit{Efficiently Computable:} For any $g_1 \in \mathbb{G}_1$, $g_2 \in \mathbb{G}_2$, there exists an efficient algorithm to compute $e(g_1, g_2)$.
        \item \textit{Non-degenerate:} The generators $g_1$ and $g_2$ should satisfy $e(g_1, g_2) \neq 1$.
    \end{itemize}

\subsection{Decisional Bilinear Diffie-Hellman Assumption}
    The decisional bilinear Diffie-Hellman (DBDH) is a computational hardness assumption about the Bilinear Diffie-Hellman (BDH) problem. The basis for security of our presented architecture is formed by the hardness of DBDH. The DBDH problem is defined as follows.
    
    In a group $\mathbb{G}_1$ and $\mathbb{G}_2$ of prime order $p$, let $a, b, c \in \mathbb{Z}_p$ be choosen at random, $e : \mathbb{G}_1 \times \mathbb{G}_2 \rightarrow \mathbb{G}_3$ be the bilinear mapping, $g_1$ and $g_2$ be generators of $\mathbb{G}_1$ and $\mathbb{G}_2$, respectively. When given $(g_1, g_2, g_1^a, g_1^c, g_2^a,g_2^b)$, the adversary $\mathcal{A}$ must be able to distinguish $e(g_1, g_2)^{abc} \in \mathbb{G}_3$ from a random element $\mathcal{R} \in \mathbb{G}_3$
    
\subsection{Cipher Policy Attribute Based SignCryption}
    The attribute-based signcryption (ABSC) is a logical combination of the attribute-based signature (ABS) along with the attribute-based encryption (ABE) in order to take the advantages of both digital signature and encryption. It adopts ABS to provide anonymous authentication, whereas, ABE to ensure data confidentiality as well as fine-grained access control. In fact, in this work, to support our blockchain-assisted architecture and ensure targeted distribution of messages, we consider an ABE-based multicast variant, i.e., ABSC scheme presented in \cite{hu2015attribute}, which is based on \cite{bethencourt2007ciphertext}. This ABSC scheme is based on cipher-policy, thus, it is referred as CP-ABSC. The CP-ABSC encrypts and signs the message with an access policy.
    
    A typical CP-ABSC scheme basically has four algorithms as presented below.

    \begin{itemize}
    \item $ABSC.Setup(1^\lambda) \rightarrow (pk, mk)$: The setup algorithm accepts as input $\lambda$ which is a security parameter. After that, it returns public parameters $pk$ and a master key $mk$.
     
    \item $ABSC.KeyGen(pk, mk, \mathcal{S}) \rightarrow (sk, key_{sign}, key_{ver})$: Given $pk$, $mk$, and a set of attributes $\mathcal{S}$, this algorithm generates a secret key $sk$, a signing key $key_{sign}$ as well as a verification key $key_{ver}$ as outputs.
    
    \item $ABSC.SignCrypt(pk, M, \mathcal{T}, key_{sign}) \rightarrow \mathcal{ST}$: The signcryption algorithm takes as input the public parameter $pk$, the plain-text $M$, the access tree $\mathcal{T}$, and the signing key $key_{sign}$. Then, the algorithm outputs a signed ciphertext $\mathcal{ST}$ of the plain-text with respect to the $\mathcal{T}$.
    
    \item $ABSC.DeSignCrypt(ST, sk, \mathcal{S}) \rightarrow \mathcal{M}$ or $\perp$: Given the signed ciphertext $ST$, the secret key $sk$, and the set of attributes $\mathcal{S}$, the designcryption algorithm outputs $\mathcal{M}$ if $\mathcal{S}$ satisfies $\mathcal{T}$, and the error symbol $\perp$ otherwise.
    \end{itemize}
    
\subsection{Permissioned Blockchain and Consensus}
    Blockchain \cite{zong2024relac, wang2024blockchain1, liu2024aoi, yin2024enabling, mollah2020blockchainiov, mollah2020blockchain}, a tamper-proof and consensus based distributed ledger technique, consists of a chain of blocks. The security and privacy of blockchain is ensured by cryptographic techniques, consensus, and incentive mechanisms. Blockchain offers data auditability by employing authenticated blocks attached in the growing chain. In the chain, each block keeps records of financial transactions or non-financial data, and a number of blocks form a chain by including the cryptographic hash of the previous block.
    
    In our proposed architecture, we consider developing a permissioned blockchain. Permissioned blockchain concept has been introduced for such applications where write permission is not necessary for end-users, and a number of trusted pre-selected users are permissioned to validate and create new blocks. Besides, Proof of Authority (PoA) \cite{misc1, misc2} is adopted as the consensus mechanism in our architecture, rather than Proof of Work, Proof of Stake, or Practical Byzantine Fault Tolerance, where trusted and pre-defined participating nodes are selected to maintain the blockchain.
    
    In the same connection, our permissioned blockchain network consists of two kinds of entities, such as PoA validators and user devices. The validators which can be chosen among the service providers or vendors over time which are restricted to validate a message (also, referred as transaction) to create a new block, maintain the consistency of the chain, and keep updated the entire blockchain in its storage. Once a validator validates a received message, the message will be recorded into a new block and appended in the main blockchain. On the other hand, the user devices including EDs and SDs are public nodes, and they are permissioned to read only. Moreover, before appending the newly validated block to the end of the current chain, these user devices are able to check the integrity of the blockchain accordingly.
    
\section{The Proposed STarEdgeChain}
    In this section, we will discuss the construction of STarEdgeChain, the proposed secure targeted message dissemination approach considering four phases: system setup, entity registration, block creation \& validation, and message dissemination.
    
\subsection{System Initialization}
    In this phase, TA runs the $ABSC.Setup$ algorithm to compute the public parameters. In general, TA will be engaged only in the system initialization and entity registration phases, and it will remain offline once finished the tasks for those phases. First, for a specific service provider having pseudo identity $id$, the algorithm generates three cyclic groups $\mathbb{G}_1$, $\mathbb{G}_2$, and $\mathbb{G}_3$ of prime order $p$ with the generators $g_1$ and $g_2$ for $\mathbb{G}_1$ and $\mathbb{G}_2$, respectively, according to the security parameter $\lambda$, and an efficiently commutable bilinear mapping $e : \mathbb{G}_1 \times \mathbb{G}_2 \rightarrow \mathbb{G}_3$. 
    
    Second, it chooses $\alpha, \beta \in \mathbb{Z}_p$, which are two random exponents. The master key is given by $mk = (\beta, {g}_2^{\alpha})$. Then, it also chooses hash functions $\mathcal{H}_1 : \{0, 1\}^* \rightarrow \{0, 1\}^\lambda$ and $\mathcal{H}_2 : \{0, 1\}^* \rightarrow \mathbb{Z}_p$. Next, it computes $h = g_1^{\beta}$ and $t = e(g_1, g_2)^{\alpha}$. Last, TA releases the public parameters to all the involved entities of the blockchain network as follows.
    
    \begin{equation}
    pk = (p, {\mathbb{G}}_1, {\mathbb{G}}_2, \mathcal{H}_2, g_1, g_2, h, t).
    \end{equation}

\subsection{Entity Registration}
    When the SP joins the blockchain network, they have to go through a registration phase and TA then authenticates both participants. Once authenticated, TA runs $ABSC.KeyGen$ algorithm which first chooses random values $r_{enc}, r_{sign} \in {\mathbb{Z}}_{p}$ and computes $D_{enc} = g_2^{\frac{{\alpha + r_{enc}}}{\beta}}$, $key_{sign} = g_2^{\frac{{\alpha + r_{sign}}}{\beta}}$, and $key_{ver} = g_2^{r_{sign}}$. TA also computes $D_j = g_2^{r_{enc}} \cdot g_2^{(\mathcal{H}_{2}(j) \cdot r_j}$ and $D_j^\prime = g_1^{r_j}$ after picking another value $r_j \in \mathbb{Z}_p$ randomly for each attribute $j \in \mathcal{S}$. Here, $\mathcal{S}$ represents the set of attributes. Next, TA returns the asymmetric secret key given by
    
    \begin{equation}
        sk = (D_{enc}, \forall j \in \mathcal{S} : D_j, D_j^{\prime}). 
    \end{equation}	
    
    Here, $key_{sign}$, $key_{ver}$, and $sk$  are considered as signing key, verification key, and decryption key, respectively. Finally, TA assigns $key_{sign}$ to the participant SP and $sk$ to the SD of attribute set $\mathcal{S}$ through a secure communications channel. Besides, the SD which are not connected with TA, the $sk$ may be pre-saved in the SD securely.
    
\subsection{Block Creation and Validation}
    To send a message $msg$ to the blockchain network for the smart devices SD, the registered service providers SP generates a random symmetric secret key $key_{sym}$ and encrypts the $msg$ with $key_{sym}$ which results the cipher-text $\mathcal{C}_{msg}$. After that, the SP defines a tree access structure $\mathcal{T}$ representing a set of SD satisfying the access policy, i.e., to control the SD's access to the encrypted message. Each non-leaf node of access tree $\mathcal{T}$ describes a threshold gate, followed by children and threshold value, and each leaf node represents an attribute. Let $num_x$ = number of children of $x$, $k_x$ = threshold value, each interior node x has such two parameters such as $num_x$ and $k_x$, and $1 \leq k_x \leq num_x$. For $k_x = 1$ and $k_x = num_x$, the threshold gates are OR and AND, respectively, and each leaf node of $\mathcal{T}$ is represented by an attribute and $k_x = 1$. Let also define the functions index(x): indicates the order of the nodes, helps to set unique values, and the children are indicated from 1 to num and attribute(x): the attributes of leaf node $x$. Once defined $\mathcal{T}$, the SP signcrypts the $key_{sym}$ under $\mathcal{T}$ by following detailed procedures.
    
    First, the service provider runs the $ABSC.SignCrypt$ algorithm, and the algorithm picks a polynomial $q_x$ for each node $x$ including its leaves in $\mathcal{T}$. These polynomials are picked in a top-down way which starts from the root node $\mathcal{R}$. For each node $x$ in $\mathcal{T}$, it sets the degree $d_x = k_x - 1$, where $k_x$ refers to the threshold value. And, starting with $\mathcal{R}$, it selects a value $s \in \mathbb{Z}_p$ randomly and also assigns $q_R(0) = s$. Then, it picks $d_R$ values from $\mathbb{Z}_p$ randomly to define $q_R$ completely. On the other hand, for any other node $x$, it assigns $q_x(0) = q_{par.(x)} (index(x))$ and picks values $d_x$ from $\mathbb{Z}_p$ randomly to define $q_x$ completely.
    
    Let $\mathcal{Y}$ represents the set of leaf nodes in the access tree structure $\mathcal{T}$. The SP picks a value $\zeta \in \mathbb{Z}_p$ randomly and execute $\tilde{\mathcal{C}} = key_{sym} \oplus t^s$, $\mathcal{C} = h^s$, $\forall y \in \mathcal{Y} : \mathcal{C}_y = g_1^{q_y(0)}$, $\mathcal{C}_y^{\prime} = g_1^{(\mathcal{H}_2(attribute(y)) \cdot q_y(0))}$, $\delta = e(\mathcal{C}, g_2)^{\zeta}$, $\pi = \mathcal{H}_1(msg) + \mathcal{H}_2(\delta)$, $w = g_1^s$, and $\psi = g_2^{\zeta} \cdot (key_{sign})^{\pi}$.
    
    By doing so, the signcrypted cipher-text $\mathcal{ST}$ under the $\mathcal{T}$ is formed as:
    
    \begin{equation}
        \mathcal{ST} = (\mathcal{T}, \tilde{\mathcal{C}}, \mathcal{C}, \forall y \in \mathcal{Y} : \mathcal{C}_y, \mathcal{C}_y^{\prime}; w, \pi, \psi)
    \end{equation}
    
    However, given service provider's public key $pk$ as well as pseudo identity $id$, hash of block $h_2$, and signcrypted cipher-text $\mathcal{ST}$ including service provider’s signature ${\pi}$, the record will form as follow and it will send to the blockchain network for validation.
    \begin{equation}
        \mathcal{R} = ({pk}, {id}, {h_2}, \mathcal{ST})
    \end{equation}
    
    In particular, we consider all registered service providers as a set of authority candidates for PoA validation in STarEdgeChain. Among them, one (self) or multiple pre-authenticated authorities can be responsible for making each new block of records which will eventually be attached with the blockchain. And, there will be no validation rewards. The process will be as follows.
    
    \begin{itemize}
    \item A genesis block, the first block, will be created once the permissioned chain will be initiated.
    
    \item Let the time is allocated into a number of intervals $t = \{t_1, t_2, t_3, t_4, ..., t_n\}$ so that the permissioned chain is able to be affixed with single blocks in each time intervals.
    
    \item In case of multiple authorities, within the time interval, one authority will be selected as a leader to collect the record $\mathcal{R}$ and validate it before passing it to others candidate authorities. Once validated by a pre-defined specific number of authorities, the new block will consist of record $\mathcal{R}$ and blockheader (block index, hash of previous block, pseudo identity ${id}_i$, and timestamp).
    \end{itemize}
    
\begin{figure*} [!t]
	\includegraphics[width=\linewidth]{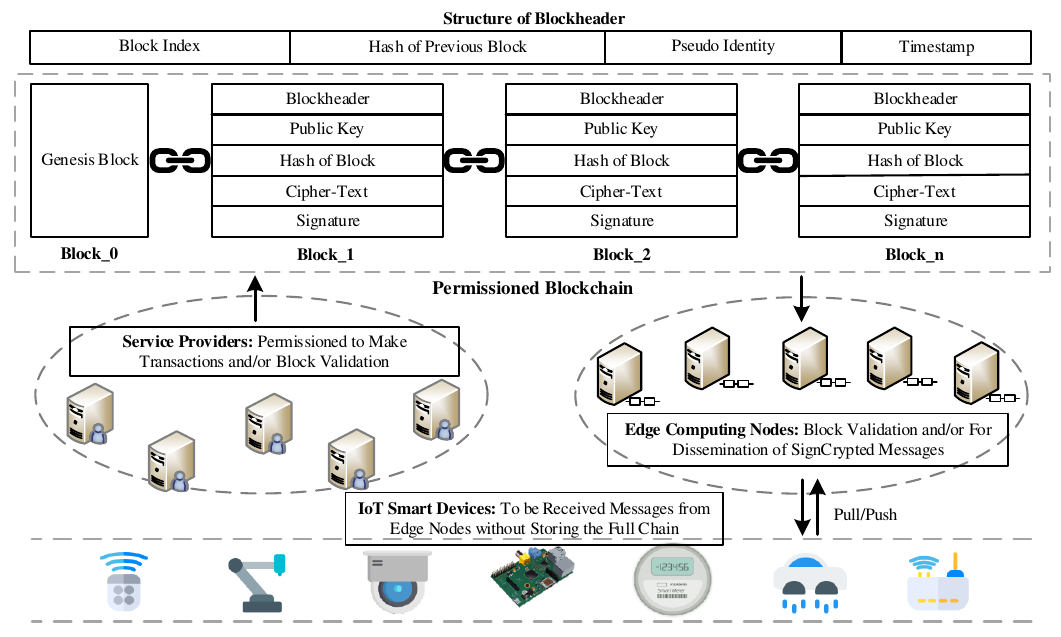}
	\centering
	\caption{The construction of permissioned blockchain presented in this work.}
	\label{fig: contructionofchain}
\end{figure*}

    The construction of the permissioned blockchain is illustrated in Fig. \ref{fig: contructionofchain}.

\subsection{Message Dissemination}
    Once received a new block, the edge computing may consider  forwarding the blockchain header to SD so that SD can decide to do a pull request for downloading the signed ciphertext $\mathcal{ST}$ and the signature $\pi$ from the specific blocks. On the other hand, the messages can also be forwarded to SD from ED (push-based approach). In this case, if the SDs are not connected with the TA, the SD owners can pre-load the $sk$ in their devices. However, after receiving the $\mathcal{ST}$ and $\pi$ from the blockchain, the SD first decrypts to recover the symmetric key $key_{sym}$ and subsequently $msg$, then checks the integrity of $msg$, and finally verifies the signature as follows.
    
    First, the SD executes the $ABSC.DesignCryption$ algorithm. A recursive algorithm $DecryptNode(\mathcal{ST}, sk, x)$ is considered which takes $pk$, $\mathcal{ST}$, and $sk$ as inputs. Here, the $sk$ possesses the set of attributes and a node $x$ from the access tree $\mathcal{T}$.
    
    For the case of $x$ is a leaf-node, let $i = attribute(x)$. And, for this case, the algorithm will return as

\begin{equation}
        DecryptNode(\mathcal{ST}, sk, x) = 
        \begin{cases}
        e(g_1, g_2)^{r_{enc}q_x(0)}, & i \in \mathcal{S} \\
        \perp, & \text{otherwise}.
        \end{cases}
\end{equation}
    
    On the other hand, if $x$ is a non-leaf node, the algorithm begins by calling $DecryptNode(\mathcal{ST}, sk, z)$ for each child node $z$ of node $x$, and the output will be stored as $\mathcal{F}_z$. Let $\mathcal{S}_x$ be a set of child nodes of $x$ with an arbitrary size $k_x$ and have $\mathcal{F}_z \neq \perp$ for all child nodes $z$. If the aforesaid set $\mathcal{S}_x$ exists, the algorithm executes as below to obtain $\mathcal{F}_z$, where $i_z = index(z)$ and $\mathcal{S}_x^{\prime} = \{ index(z) \Vert z \in \mathcal{S}_x \}$
    
    \begin{equation}
        \mathcal{F}_x = \prod_{z \in \mathcal{S}_x} \mathcal{F}_z^{{\Delta}_{{{i}_{z}}, \mathcal{S}_x^{\prime}}(0)} = e(g_1, g_2)^{r_{enc}q_x(0)}
    \end{equation}
    
    After that, the $ABSC.DesignCryption$ algorithm calls the function of the root node $r$ of the access tree. The SD is able to recover the symmetric secret key subject to holding the set of attributes $\mathcal{S}$. Once satisfied, let set $DecryptNode(\mathcal{ST}, sk, r) = A = e(g_1, g_2)^{r_{enc}q_r(0)} = e(g_1, g_2)^{r_{enc}s}$. 
    
    The decrypted symmetric key $key_{sym}^{\prime}$ will be obtained by computing as
    
    \begin{equation}
     \frac{\tilde{\mathcal{C}}}{\frac{e(\mathcal{C}, D_{enc})}{A}} = \frac{key_{sym}^{\prime} e(g_1, g_2)^{\alpha s}}{e(g_1, g_2)^{\alpha s}} = key_{sym}^{\prime}
    \end{equation}
    
    Second, the SD also computes $\delta^{\prime} = \frac{e(\mathcal{C}, \psi)}{{(e(w, key_{ver}) \cdot \tilde{A})}^{\pi}}$, where $\tilde{A} = \frac{e(\mathcal{C}, D_{enc})}{A}$.
    Once obtained the $key_{sym}^{\prime}$, the SD utilizes it to decrypt the message as $msg^{\prime}$. Next, it calculates $\mathcal{H}_1(msg^{\prime}) + \mathcal{H}_2(\delta^{\prime})$, if it will be equal to $\pi$, the SD is confirmed that the $msg$ has not been modified.
    
    A discussion on the proposed permissioned blockchain architecture is as follows in terms of the following two point of views.
    
    \textit{(i) Growing Chain:} One of the key challenges of blockchain is growing the size of the chain over time since blockchain maintains a growing immutable ledger. In fact, this incurs extra storage burden of the computing nodes who are maintaining the blockchain. However, unlike the transactions in cryptocurrency applications, there might be no relationship with one transaction to other transactions due to being non-cryptocurrency application, and thus, the old transactions might be useless after some time. Moreover, we do not employ the Merkle tree data structure which is used to make a summary of all transactions. In the same connection, the permissioned nodes in the blockchain network will have flexibility to close down the chain in order to reduce the storage burden.
    
    \textit{(ii) Consensus Mechanism:} Most of the public blockchains employ proof of work (PoW), proof of stake (PoS), etc. mechanisms to validate the transactions. In particular, PoW requires to solve some puzzle which usually requires a tremendous amount of computational resources. Thus, for the purpose of reaching consensus, our architecture utilizes proof of authority (PoA) mechanism. Besides, PoS is not applicable in our application since our message dissemination application is a non-cryptocurrency application, in which we do not have any currency or balance as a stake.

\section{Security and Performance Analysis}
    In this section, we present the analysis of our introduced architecture from the security and privacy point of view and the implementation details as well as measurement of computational overheads.
    
\subsection{Security and Privacy Analysis}
    In the following, we describe how our proposed architecture will fulfill the security and privacy goals.
        
    \textit{Privacy Preservation:} A passive adversary $\mathcal{A}$ may be curious to retrieve the details of a message to exploit known as well as zero-day vulnerabilities to launch future attacks. To do so, the adversary may go for monitoring the sender of records $\mathcal{R} = ({pk}, {id}, {h_2}, \mathcal{ST})$, investigating the block contents, and/or analyzing the pull requests (at edge computing device or cache repository sides) from the smart devices in order to gather the knowledge of messages, such as service provider's information, etc. However, in our proposed architecture, we have used pseudo identities for all blockchain network entities instead of real identities, and we have not used any sensitive attribute values while forming the access tree $\mathcal{T}$. Hence, it might be impossible for the adversary to infer the details of messages unless the adversary will be able to access the trusted authority and map the pseudo identities and their respective real identities. Therefore, we claim that our proposed architecture can protect the privacy of the engaged entities in the blockchain network.
    
    \textit{Message Authenticity:} According to the cryptographic scheme employed in our proposed architecture, the legitimate message is sign-crypted referred as $\mathcal{ST}$, whereas the TA will make $key_{ver}$ for each service providers as public parameter so that any entity including block validators and smart devices will be able to verify the authenticity of the messages. According to the scheme, it may be infeasible for any curious insiders and outsiders to reconstruct $key_{ver}$. This is because according to the system initialization and entity registration subsections of section IV, the $key_{ver}$ comes by calculating the $g_2^{r_{sign}}$, and here, $g_2$ and $r_{sign}$ are randomly generated by the TA. That is, we confirm that our proposed architecture obtains message authenticity property.
    
    \textit{Confidentiality of Messages:} The proposed architecture will obtain the confidentiality of messages which are sent by the service providers to the blockchain network to reach particular smart devices. Such confidentiality is mainly relied on a cipher policy attribute based signcryption scheme. However, to recover the messages, the smart devices have to retrieve the symmetric key first. As shown in Eqn. 7, in order to retrieve the symmetric key successfully from received $\mathcal{ST}$, the adversary $\mathcal{A}$ has to reconstruct $e(g_1, g_2)^{\alpha s}$. In fact, to retrieve the component $A$ which includes secret key $sk$ and set of attributes according to Eqn. 5. Consequently, only IoT smart devices which will match the list of attributes and have the secret key $sk$ can retrieve the messages requested from the blockchain or received from the edge devices. In the same connection, it’s applicable for all curious blockchain nodes. Thus, for any illegitimate entities including curious blockchain nodes and malicious entities could not have sufficient information to retrieve the messages. By this way, each message confidentiality against any curious nodes is assured.
    
    \textit{Resilience to Integrity Attacks:} The proposed architecture obtains the integrity of messages in two ways. First, as shown in Eqn. 3, the sign-crypted cipher-text $\mathcal{ST}$ consists of three signature parameters, i.e., $\delta$, $\pi$, and $\psi$. Among them, the $\pi$ was calculated by making the hash of the message. Second one is more related to blockchain's tamper-proof feature. And, according to the blockchain structure, every blockheader keeps the hash of the previous block to make a digitally connected chain. In fact, due to aforementioned reasons, if the messages are compromised by any integrity type of attacks, such as replacing the valid message with a malicious one, it can be detected. Thus, the proposed architecture is resilient to the integrity attacks.
    
    \textit{Access Control:} The access control functionality between the service providers and targeted smart devices can be inherently realized in our proposed architecture with the help of utilized cryptographic scheme. In other words, the service providers can pre-define the set of attributes to provide access control according to block creation and validation subsection of section IV. Thus, we do confirm that our proposed design can obtain access control.
    
    \textit{Freshness of Messages:} Since each block creation time is saved as timestamp in the blockheader part of block, our architecture can resist any attempt by a passive adversary $\mathcal{A}$ to send an old message. The aim of sending old messages is to obtain specific messages, which may have known vulnerabilities or intentionally falsify the smart devices. With the timestamp, the smart devices can verify the freshness of messages by themselves.
    
    \textit{Resistance to Availability Attacks:} Resistance against availability attacks is crucial for IoT since many smart devices are there in the system. In the proposed architecture, the availability of attacks including denial of services (DoS), distributed DoS (DDoS) or single point of failure (SPoF) might be avoided with the help of blockchain. This is because employing blockchain’s decentralized feature, i.e., blockchain is usually maintained by multiple nodes. Although, we have used a trusted authority, basically for the purpose of registration and key generations, is different than blockchain storage nodes. Therefore, we have claimed that our proposed architecture is resistant against availability types of attacks.
    
    \textit{Traceability:} Although we have used pseudo identities instead of real ones, in case of any malicious activities by the valid and registered entities through insider attacks or compromised by external adversaries can be traceable. For this, the trusted authority, responsible for entity registration, will support to find out the specific nodes from his database by mapping the pseudo identities and respective real identities. Thus, our proposed design can achieve traceability of misbehaving nodes.

\subsection{Implementation and Performance Evaluation}
    In this subsection, we discuss a software prototype implementation of our proposed permissioned blockchain as presented in section IV. One of the aims of this implementation is to observe the computational overheads required in all involved entities which will be presented as follows.

\begin{figure*} [!t]
	\includegraphics[width=.85\linewidth]{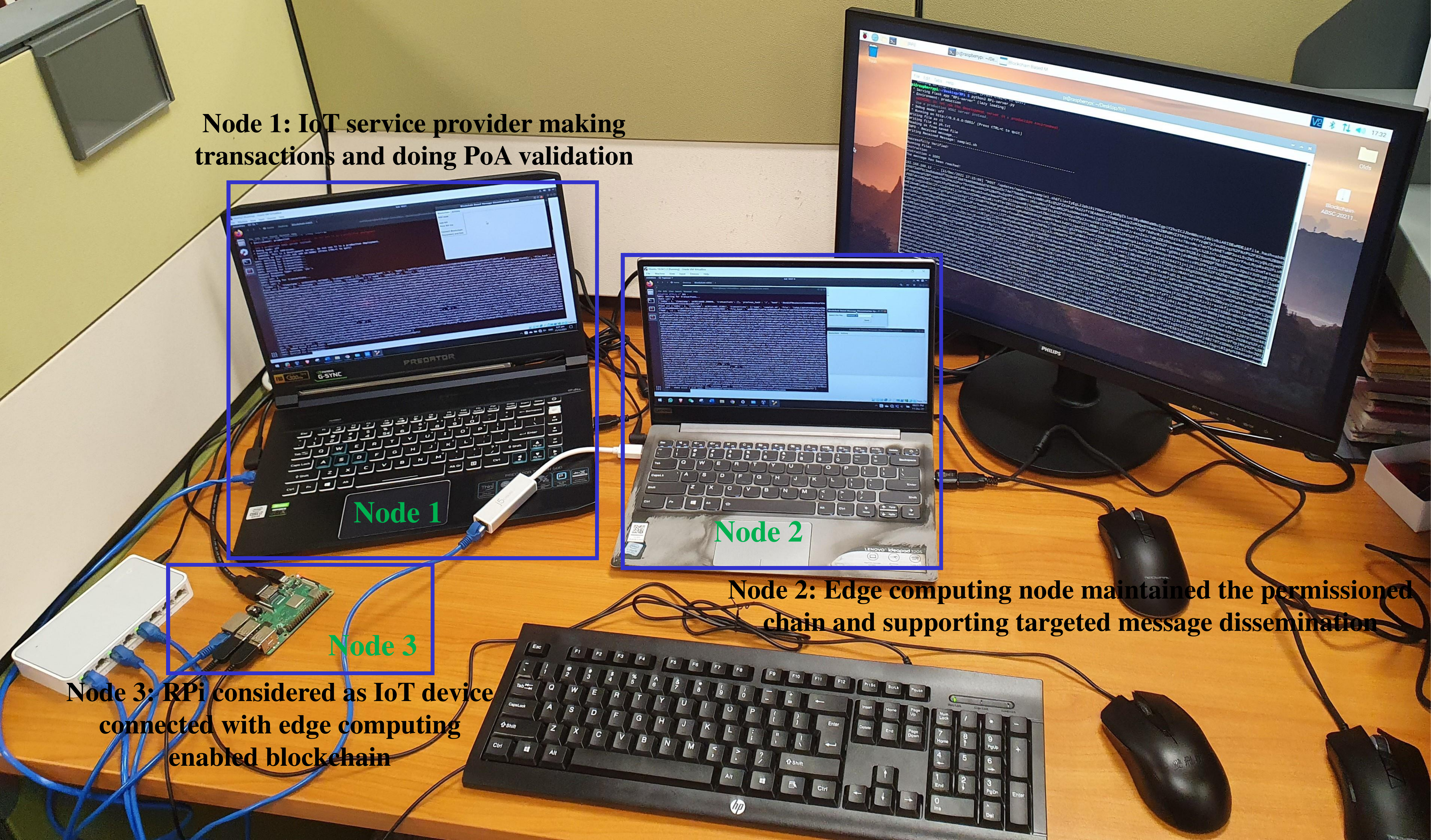}
        \centering
	\caption{The experimental setup of our proposed architecture with permissioned blockchain.}
    \label{fig: experiment}
\end{figure*}
    
    \textit{Implementation Environment:} The service providers, PoA validators, and edge computing nodes commonly have high computational devices relatively to IoT smart devices. Thus, as presented in Fig. \ref{fig: experiment}, we have taken two laptop computer, representing as two host machines (Acer Predator Triton 500 and Lenovo IdeaPad 320S) as service providers, PoA validators, and edge computing nodes, whereas, one Raspberry Pi device (RPi 3 Model B+) as IoT smart device. In the host machines, we have configured Oracle virtual box (https://www.virtualbox.org/) and installed Ubuntu 18.04 with one core, RAM 4GB as virtual machines. In fact, we have considered the aforementioned virtual machines and one smart device as node 1, 2, and 3, respectively. 
    
    Besides, we have used Python 3.6 version for developing the software prototype. After that, we have used the bridge network feature from Oracle virtual box and made a local network among the host machines, virtual machines, and the RPi device for the purpose of on-chain implementation. We have also developed a graphical user interface (GUI) for convenience operation. Specifically, in Fig. \ref{fig: experiment}, the node 1 represents a service provider making transactions and doing PoA self-validation of transactions to add in the chain. On the other hand, node 2 shows an edge computing node connected with a blockchain network and supported to perform targeted message dissemination to the RPi device considered as IoT smart device in node 3. 

    The codes and implementation instructions are publicly available to facilitate the reproducibility of this work at Github\footnote{\url{https://github.com/mbaqer/Blockchain-IoT}}. The Table \ref{tab: setup_summary} presents a summery of the computing nodes, assigned IP addresses \& ports, used libraries, and other configurations. In particular, we have utilized Flask and Tkinter to develop a web based application and GUI, respectively, whereas the Charm-Crypto and hashlib are cryptography libraries, which are discussed in details in the next paragraph.
    
\begin{table}[ht]
    \centering
    \caption{Summary of setup, platform, and configuration employed in the experiments.}
    \label{tab: setup_summary}
\begin{tabular}{m{2.5cm}|m{5.5cm}}
    
    \hline \hline

    \textbf{Components}   &   \textbf{Details} \\
    \hline

    Node 1  &    Intel Core i7-10th generation, RAM 32GB, and Microsoft Windows 10 operating system)   \\
    \hline
    
    Node 2  &    Intel Core i5-8th generation, RAM 8GB, and Microsoft Windows 10 operating system)   \\
    \hline
    
    Node 3  &  ARM Cortex-A53 (ARMv8) 1.4GHz quad-core processor, 1GB RAM, and Raspberry Pi OS) \\
    \hline
    
    Libraries    & Flask, Tkinter, Charm-Crypto, hashlib   \\
    \hline
    
    IP and Port addresses    & Node 1 VM1: 192.168.100.10:5000, \newline 
    Node 1 Host1: 192.168.100.11 \newline 
    Node 2 VM2: 192.168.100.12:5000 \newline 
    Node 2 Host2: 192.168.100.13 \newline 
    Node 3 RPi: 192.168.100.14:5001 \\
    \hline
    
    Pre-set PoA consensus time     & 15 $sec$  \\ 
    \hline

\end{tabular}
\end{table}

\begin{figure*} [ht!]
\begin{subfigure}[b]{0.30\textwidth}
	\centering
	\includegraphics[width=5.8cm]{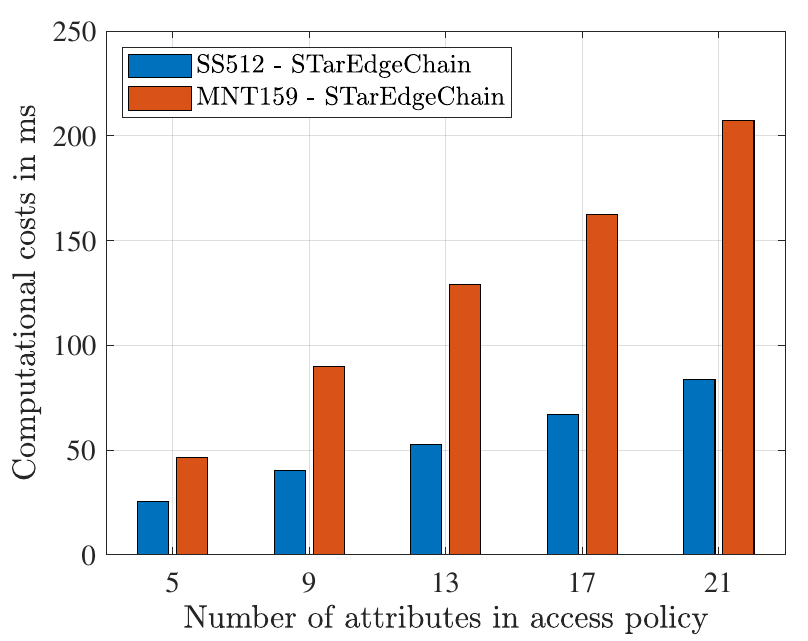}
	\caption{\scriptsize Setup and key generation at Trusted Authority end}
\end{subfigure}
    \hfill
\begin{subfigure}[b]{0.30\textwidth}
	\includegraphics[width=5.8cm]{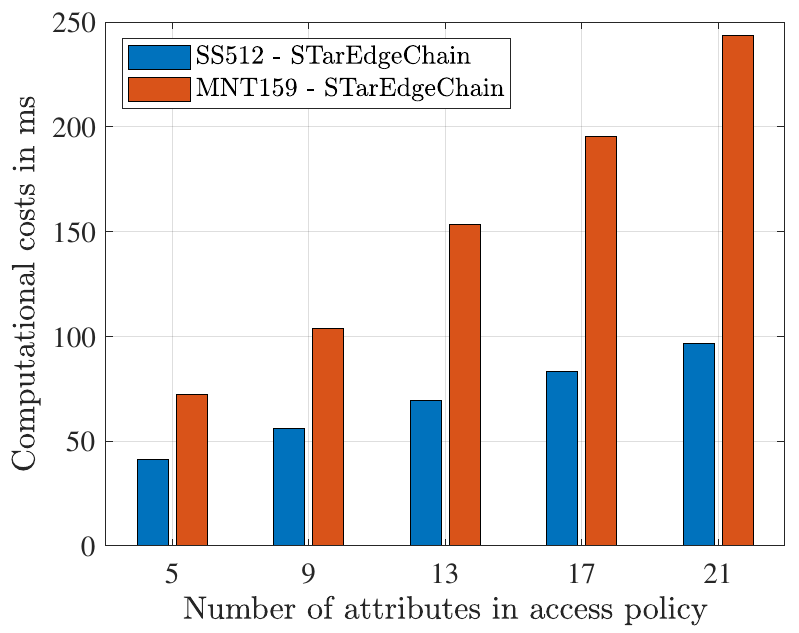}
	\centering
	\caption{\scriptsize SignCryption on block creation \& validation phase}
\end{subfigure}
	\hfill
\begin{subfigure}[b]{0.30\textwidth}
    \centering
	\includegraphics[width=5.8cm]{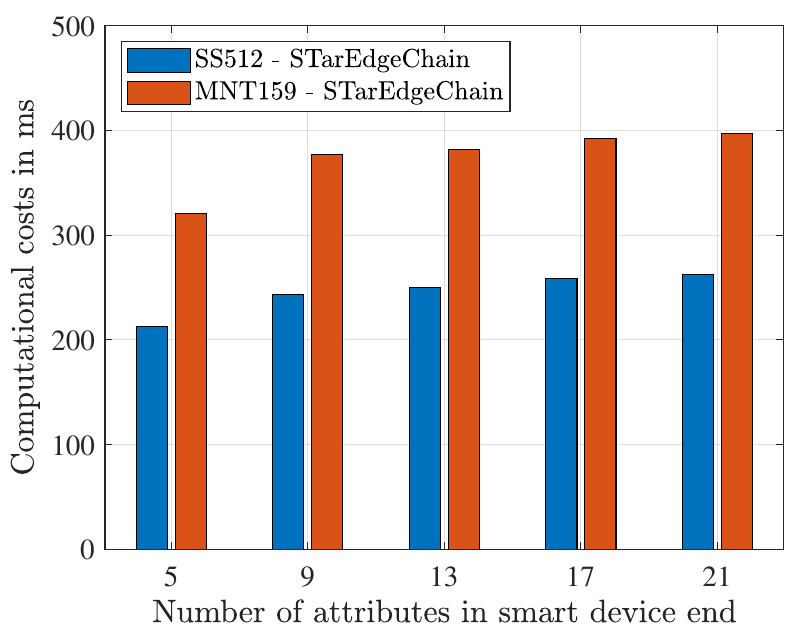}
	\caption{\scriptsize DeSignCryption after message dissemination}
\end{subfigure}

\vspace{4mm}

\begin{subfigure}[b]{0.30\textwidth}
    \centering
	\includegraphics[width=5.8cm]{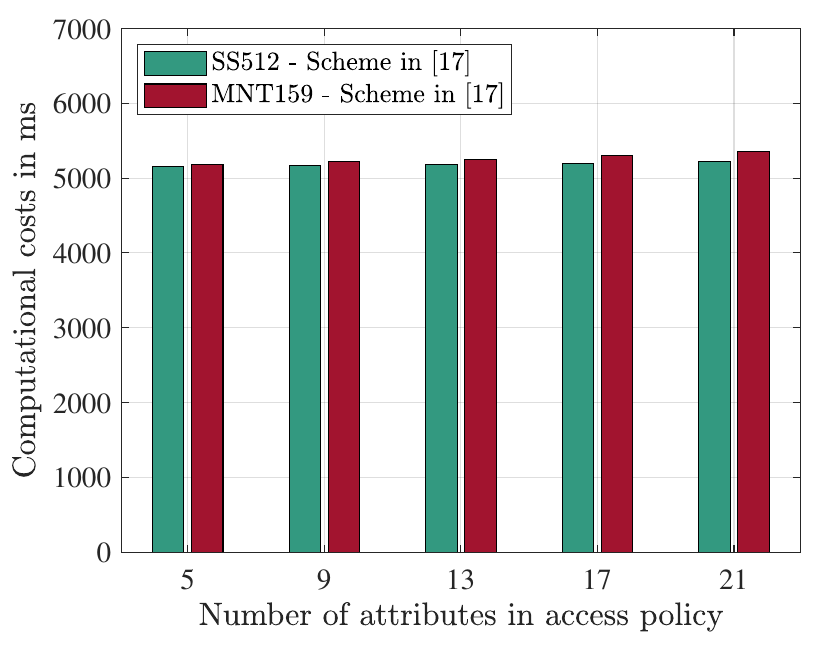}
	\caption{\scriptsize Setup and key generation at Trusted Authority end implemented by \cite{ambrosin2014updaticator}}
\end{subfigure}
	\hfill
\begin{subfigure}[b]{0.30\textwidth}
    \centering
	\includegraphics[width=5.8cm]{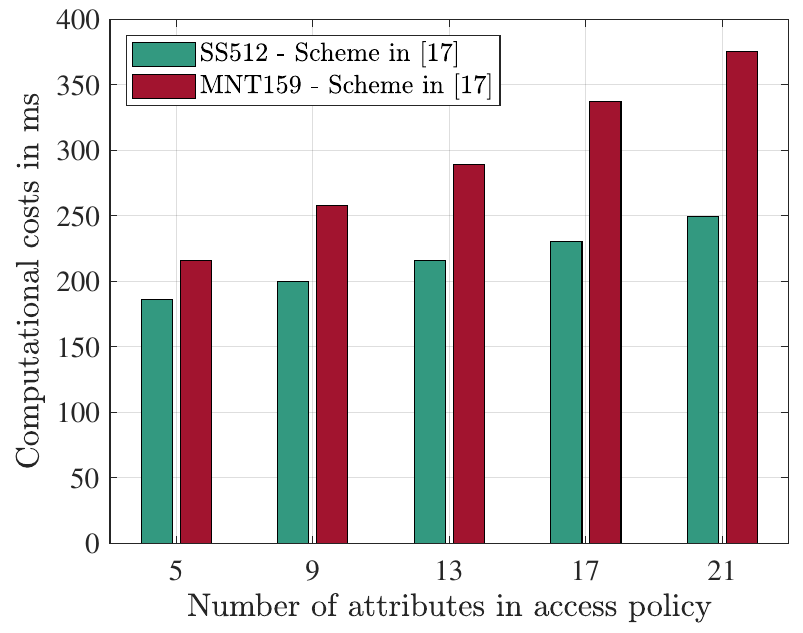}
	\caption{\scriptsize Sign and encryption implemented by scheme in \cite{ambrosin2014updaticator}}
\end{subfigure}
	\hfill
\begin{subfigure}[b]{0.30\textwidth}
    \centering
	\includegraphics[width=5.8cm]{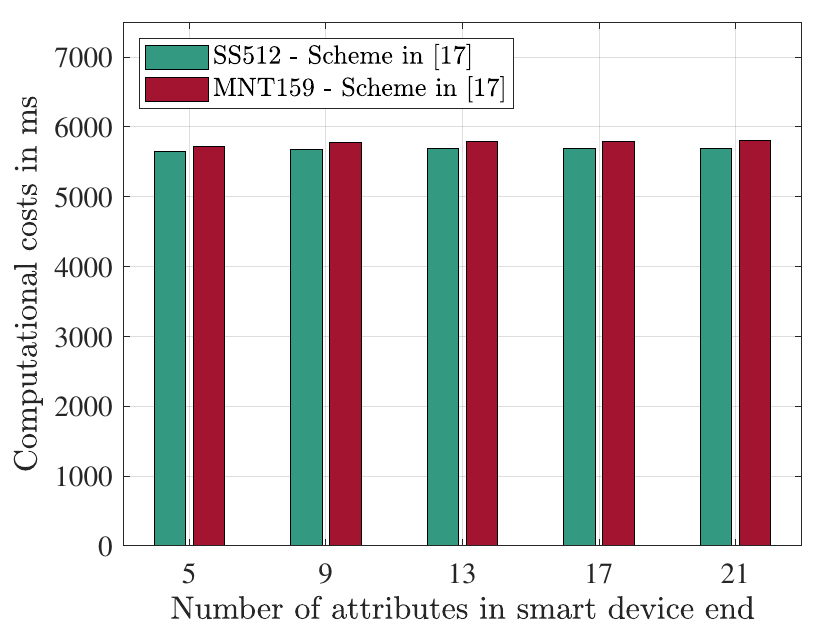}
	\caption{\scriptsize Verify and decryption implemented by scheme in \cite{ambrosin2014updaticator}}
\end{subfigure}
	\caption{Performance analysis in terms of time consumption with different access policies and attribute sizes}
	\label{fig: CP-ABSC_Results}
\end{figure*}

    \textit{Analysis of Computational Costs:} In our experiments, advanced cryptographic schemes are employed such as CP-ABSC \cite{hu2015attribute}, SHA-256 as well as AES (CBC mode) to fulfill the security and privacy requirements in our proposed architecture. Especially, to implement cryptographic parts, we have utilized a pairing based cryptography library referred as Charm \cite{akinyele2013charm}. In particular, we have implemented both SS512 (symmetric elliptic curve) and MNT159 (asymmetric elliptic curve)  for all pairing operations in our experiments. Besides, we have considered a shell script (.sh file) as an executable file with size 1 MB. And, we have used SHA-256 for hashing, and in experiment, we have utilized hashlib \cite{misc3}. On the other hand, we have also compared the cryptographic performances with another work presented in \cite{ambrosin2014updaticator}, in which the authors employ a multicast ABE variant, i.e., the scheme also does not consider individual entity revocation. Basically, we have chosen this work for the purpose of comparison since both ours and their work’s utilized schemes are commonly based on \cite{bethencourt2007ciphertext}. Specifically, in their utilized scheme, the authors employ RSA and AES (CBC mode) with key size of 4096 and 256 bits, respectively, along with the CP-ABE scheme. In our experiment, we considered calculating the time complexities of the sign/verify and encryption/decryption together. However, the results of all cryptographic performances from the experiments are presented in Fig. \ref{fig: CP-ABSC_Results} as summary, where subfigures \ref{fig: CP-ABSC_Results}(a), \ref{fig: CP-ABSC_Results}(b), \ref{fig: CP-ABSC_Results}(c), \ref{fig: CP-ABSC_Results}(d), \ref{fig: CP-ABSC_Results}(e), and \ref{fig: CP-ABSC_Results}(f) are referred as different operations, such as setup \& key generation, signcryption, designcryption, signing \& encryption, and verification \& decryption. Here, all operations have been measured 5 times to calculate the average value. We have set different size of attributes ranging from 2 to 19, and from the results, we can see the computational costs with different sizes at attributes.

    Overall, the computational costs introduced by the cryptographic processing in our developed permissioned blockchain are very modest for the lightweight IoT smart devices. Usually, the lightweight IoT smart device nodes have very limited storage to keep the entire blockchain and update it over time. Therefore, the interesting point is that in our permissioned blockchain, the lightweight nodes do not require to store the entire blockchain, rather the nodes will be relied on edge computing nodes. This shows the advantages of employing this work into wearable devices (smartwatches and fitness trackers), smart home devices (smart speakers, thermostats, security cameras, lights, and appliances), and healthcare IoT gadgets (blood pressure monitors, glucose meters, and ECG devices like personal health monitoring devices), just to name a few.

\section{Related Works}
    In this section, we will summarize and investigate the most relevant works on typical message dissemination, attribute-based encryption (ABE) for message dissemination, and blockchain assisted message dissemination solutions.
    
    \textit{Typical Message Dissemination:} In recent years, many solutions on typical message dissemination approaches have been introduced. For instance, the work in \cite{liu2021btmpp} focuses on secure and privacy-aware emergency vehicular message dissemination named as BTMPP. This solution utilizes bloom filter enabled private set intersection techniques to offer strong conditional privacy. Also, this proposed approach supports precise trust management simultaneously. Moreover, the authors in \cite{liu2021pptm} proposes a privacy preserving trust management approach for message dissemination in vehicular space-air-ground integrated networks, which has an aim to ensure strong applicability and robustness by employing 0-1 encoding approach. A novel low-latency cooperative security method presents in \cite{li2021confidentiality} for platoon-assisted vehicular network to ensure timely and secure vehicular data dissemination. Here, a secret key agreement scheme is employed, in which the secret keys are generated based on the correlation of vehicles radio channels. The work in \cite{falas2021modular} introduces an end-to-end fast and secure firmware update framework IoT device like embedded systems. This work mainly employed a hardware-embedded cryptographic primitive referred to as a physical unclonable function (PUF) technique. Another work presented in \cite{rabbani2020shefu} proposes a remote attestation-based firmware update for IoT. Kornaros et al. \cite{kornaros2020towards} present a secure firmware update for connected vehicles by including timestamp, version of update as well signed metadata of update.
    
    \textit{ABE in Message Dissemination:} In addition, recent works have shown that attribute-based encryption has emerged as a potential enabler to ensure confidentiality and access control requirements. As such, the work in \cite{liu2020security} presents a secure and reliable encrypted information dissemination approach for vehicular networks in multi-RSU (road-side units) settings. Here, the authors utilize two cryptographic schemes, i.e., ciphertext-policy ABE and proxy re-encryption for the purpose of providing confidentiality while broadcasting the encrypted information and ensuring to obtain the full encrypted information to the moving vehicles within their own RSU’s range, respectively. In the same connection, in \cite{xia2021policy}, a ciphertext-policy attribute-based proxy re-encryption scheme is utilized to enforce access policy accurately and timely while disseminate the data in dynamic vehicular network. Manna et al. propose an over-the-air secure software update in \cite{la2021performance} for automotive embedded platforms. A fast and secure software update distribution mechanism is introduced in \cite{ambrosin2014updaticator}. This mechanism is referred to as Updaticator, where besides ABE, a cache-enabled content distribution network service is utilized so that the update distribution can be scalable to billions of devices. Meanwhile, in \cite{tonyali2017attributes} and \cite{tonyali2017attribute}, firmware updates approaches are presented for smart meters in advanced metering infrastructure network in smart grid. Along with attribute-based encryption, in \cite{tonyali2017attributes}, a random linear network coding technique is used for the purpose of increasing the reliability of update distribution, and in \cite{tonyali2017attribute}, a multi-cast over broadcast protocol referred as broadcast alarm is utilized so that the updates can be distributed among the targeted devices.
    
    \textit{Integration of Blockchain:} In terms of blockchain integration in message dissemination, existing works mainly employ blockchain along with different cryptographic schemes. In this context, cryptographic techniques are utilized to guarantee the security and privacy related requirements to the involved entities and message contents. For example, the work in \cite{guehguih2019blockchain} constructs a public key infrastructure (PKI) based private blockchain to offer privacy-aware message dissemination in vehicular networks. Another work in \cite{cheng2019sctsc}, the authors develop a semi-decentralized attribute-based blockchain to distribute the traffic signal messages and assist traffic control among the vehicles with high availability and privacy preserving ways in IoV. Here, the authors utilize blockchain along with a ciphertext-policy ABE scheme. Besides, the authors in \cite{kumari2020blockchain} and \cite{aggarwal2019new} presents on how typical blockchain structure is able to handle data dissemination in Industrial IoT and Internet of Drones, respectively. Moreover, Baza et al. \cite{baza2019blockchain} leverage attribute-based encryption and zero-knowledge proof techniques so that the updates initially come from manufacturer can be shared among the autonomous vehicles as a peer-to-peer manner, at the same time preserving the privacy as well as being targeted distribution. In the same connection, Puggioni et al. \cite{puggioni2020towards} propose a IoT updates delivery approach by introducing a cryptocurrency and a rewarding scheme, integrated with zero-knowledge proof to address the scalability of typical update approaches. Meanwhile, a firmware update platform is proposed in \cite{hu2019autonomous} for IoT devices via hardware platform module, batch verification, digital signature, and anti-malware checking.
    
    To the best of our knowledge, the aforementioned blockchain integration with message dissemination related solutions are the most relevant to ours. However, in the following table \ref{tab: comparison}, we show how these works are different in terms of design goals.
    
\begin{table}[ht]
    \centering
    \caption{Comparison of security and privacy properties.}
    \label{tab: comparison}
    \begin{tabular}{m{2.05cm}|m{.35cm}|m{.35cm}|m{.35cm}|m{.35cm}|m{.35cm}|m{.35cm}|m{.35cm}|m{.35cm}}
    
    \hline \hline
    \textbf{Properties} & \textbf{\cite{guehguih2019blockchain}} & \textbf{\cite{cheng2019sctsc}} & \textbf{\cite{kumari2020blockchain}} & \textbf{\cite{aggarwal2019new}} & \textbf{\cite{baza2019blockchain}} & \textbf{\cite{puggioni2020towards}} & \textbf{\cite{hu2019autonomous}} & \textbf{Ours} \\
    \hline
    
    Privacy protection & \checkmark & \checkmark & \checkmark & \checkmark & \xmark & \xmark & \xmark & \checkmark \\
    \hline
    
    Authentication & \checkmark & \xmark & \checkmark & \checkmark & \checkmark & \checkmark & \checkmark & \checkmark \\
    \hline
    
    Integrity & \checkmark & \checkmark & \checkmark & \checkmark & \checkmark & \checkmark & \checkmark & \checkmark \\
    \hline
    
    Confidentiality & \xmark & \checkmark & \xmark & \xmark & \checkmark & \xmark & \xmark & \checkmark \\
    \hline
    
    Access control & \xmark & \checkmark & \xmark & \xmark & \checkmark & \xmark & \checkmark & \checkmark \\
    \hline
    
    Freshness & \xmark & \checkmark & \xmark & \xmark & \xmark & \checkmark & \xmark & \checkmark \\
    \hline
    
    Availability & \checkmark & \checkmark & \checkmark & \checkmark & \checkmark & \checkmark & \checkmark & \checkmark \\
    \hline
    
    Traceability & \checkmark & \checkmark & \checkmark & \xmark & \xmark & \xmark & \xmark & \checkmark \\
    \hline
    
    Edge security & \xmark & \xmark & \xmark & \xmark & \xmark & \xmark & \xmark & \checkmark \\
    \hline
    
    Reward payment & \xmark & \xmark & \xmark & \xmark & \checkmark & \checkmark & \xmark & \xmark \\
    \hline

\end{tabular}
\end{table}
    
\section{Conclusion and Future Works}
    Secure and privacy-aware message dissemination among targeted smart devices is one of the key concerns in the IoT system to send specific new features, safety messages, advertisements, instructions, alerts, price information, or security updates. In this paper, we have presented a security and privacy preserving targeted message dissemination solution with the use of blockchain assisted edge computing paradigm, referred as STarEdgeChain. Most importantly, we have constructed a permissioned blockchain and employed a signcryption scheme for IoT edge computing nodes. We have also discussed the security and privacy analysis of our proposed architecture and evaluated its performance by developing a software prototype. The experimental results have shown that our presented architecture can be suitable on smart devices.
    
    However, our presented architecture will be flexible enough in terms of exploring and applying other cryptographic schemes as well as blockchain consensus mechanisms which will be solely depending on the different security requirements and application scenarios. Perhaps, in future work, the research direction can be as follows: (i) investigating how to integrate offchain concepts in untrusted scenarios with zero knowledge proof, (ii) looking for mitigating the adversarial attacks against edge computing nodes, (iii) exploring how to incorporate Self-Sovereign Identity and data capsule techniques, (iv) designing and applying a signcryption scheme having decentralization registration capabilities, and/or lesser storage \& computational requirements, and (v) developing an anonymous reward payment system.

\section*{Acknowledgments}
    The authors would like to thank to the editor and reviewers for thoughtful comments and constructive suggestions toward improving the quality of this work.

\ifCLASSOPTIONcaptionsoff
  \newpage
\fi

\bibliographystyle{IEEEtran}
\bibliography{bibliography.bib}


\begin{IEEEbiography}[{\includegraphics[width=1in,height=1.25in,clip,keepaspectratio]{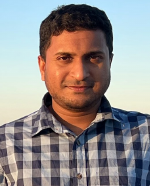}}]
    {Muhammad Baqer Mollah} is currently pursuing a PhD degree in the Department of Electrical and Computer Engineering at the University of Massachusetts Dartmouth. Before joining at University of Massachusetts Dartmouth, he has been working at School of Computer Science and Engineering in Nanyang Technological University (NTU) as a full-time Research Associate. His research interests include advanced communications, security, and edge intelligence techniques for Internet of Things (IoT) and connected vehicles. He has a M.Sc. in Computer Science and B.Sc. in Electrical and Electronic Engineering from Jahangirnagar University, Dhaka and International Islamic University Chittagong, Bangladesh, respectively.
\end{IEEEbiography}

\begin{IEEEbiography}[{\includegraphics[width=1in,height=1.25in,clip,keepaspectratio]{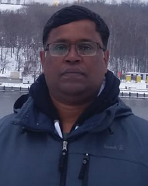}}]
    {Md Abul Kalam Azad} is currently a Professor with the department of Computer Science and Engineering at the Jahangirnagar University, Dhaka, Bangladesh. His current research interests include cloud computing, wireless networks, Internet of Things, and information security. He has a Ph.D. degree from School of Computer Engineering and Information Technology at the University of Ulsan, South Korea, M.Sc. in Information Technology (IT) from KTH Royal Institute of Technology, Sweden, and B.Sc. in Electronics and Computer Science with President Gold Medal from Jahangirnagar University, Bangladesh.
\end{IEEEbiography}

\begin{IEEEbiography}[{\includegraphics[width=1in,height=1.25in,clip,keepaspectratio]{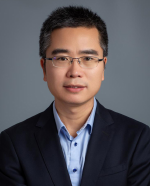}}]
    {Yinghui Zhang} is currently a Professor with the School of Cyberspace Security, Xi\textquotesingle an University of Posts and Telecommunications, Xi\textquotesingle an, China. He was a post-doctoral research fellow at Singapore Management University, Singapore. His current research includes the design and analysis of new public key cryptography schemes, cloud computing security, wireless network security, security and privacy in blockchain and machine learning, etc. He serves as an Editor/Associate Editor of several international Journals on information security. Moreover, he has served as a program committee member for several conferences. He received the Ph.D. degree in cryptography from Xidian University, Xi\textquotesingle an, China.
\end{IEEEbiography}


\end{document}